\documentclass[aps, pra, superscriptaddress, twocolumn, amsfonts, amsmath, amssymb, floatfix]{revtex4-2}
\usepackage{graphicx}
\usepackage{sidecap}
\usepackage{dcolumn}
\usepackage{bm}
\usepackage{epsfig}
\usepackage{braket}
\usepackage{color}
\usepackage{ulem}
\usepackage{amsmath}
\usepackage[toc,page]{appendix}
\usepackage[colorlinks=true, letterpaper=true, pdfstartview=FitV, linkcolor=blue, citecolor=blue, urlcolor=blue]{hyperref}
\begin{document}
\title{Parametric excitations in a harmonically trapped binary Bose-Einstein condensate}
\author{Meiling Wang}\thanks{These authors contributed equally to this work.\\}
\affiliation{Department of Physics, School of Physics and Electronic Science, East China Normal University, Shanghai 200241, China}
\author{Juan Wang}\thanks{These authors contributed equally to this work.\\}
\affiliation{Department of Physics, School of Physics and Electronic Science, East China Normal University, Shanghai 200241, China}
\author{Yan Li}
\email{yli@phy.ecnu.edu.cn}
\affiliation{Department of Physics, School of Physics and Electronic Science, East China Normal University, Shanghai 200241, China}
\affiliation{Chongqing Key Laboratory of Precision Optics, Chongqing Institute of East China Normal University, Chongqing 401120, China}
\author{Franco Dalfovo}
\email{franco.dalfovo@unitn.it}
\affiliation{Pitaevskii BEC Center, CNR-INO and Dipartimento di Fisica,
Universit\`a di Trento, via Sommarive 14, I-38123 Trento, Italy}
\author{Chunlei Qu}
\email{cqu5@stevens.edu}
\affiliation{Department of Physics, Stevens Institute of Technology, 1 Castle Point Terrace, Hoboken, NJ 07030, USA}
\affiliation{Center for Quantum Science and Engineering, Stevens Institute of Technology, 1 Castle Point Terrace, Hoboken, NJ 07030, USA}
\date{\today}
\begin{abstract}
We investigate parametric excitation and pattern formation in a harmonically trapped two-component Bose-Einstein condensate. We assume the condensate to be in the miscible phase, but near the miscible-immiscible phase transition, where total density and spin density excitations are decoupled. By periodically modulating the atomic scattering lengths, Faraday patterns can be generated in both density and spin channels. In an elongated condensate, the pattern in the spin channel corresponds to a one-dimensional standing wave with the two components exhibiting an out-of-phase density oscillation, where the modulation frequency and the oscillation period are related to the velocity of the spin sound. After the spin pattern is fully developed, the system quickly enters a nonlinear destabilization regime. For a pancake-shaped condensate, a two-dimensional Faraday pattern is generated with an interesting $l$-fold rotational symmetry. The number of nodes along the radial and angular directions increases with larger modulation frequencies. We also compare the growth rates of spin Faraday patterns generated with different modulation protocols, which are accessible to current experiments. 
\end{abstract}

\maketitle

\section{Introduction}
Parametric excitation‌ refers to a nonlinear effect caused by a periodic variation of a system parameter that produces an exponential growth of certain modes under resonance conditions. In a finite system, parametrically amplified modes can lead to pattern formation~\cite{Cross}. This is a ubiquitous phenomenon that has been studied in many research fields, such as optics~\cite{Arecchi}, cosmology~\cite{Liddle}, fluid mechanics~\cite{Faraday,Douady,Westra,Zhao}, biophysics and chemistry~\cite{Maini,Petrov}. A paradigmatic example of pattern formation was discovered in 1831 by Michael Faraday~\cite{Faraday}, who found that regular patterns can be formed on the surface of various liquids when subjected to vertical oscillatory driving. In most cases, the dynamics of parametric excitation can be understood through the Mathieu equation $\ddot{x}+\omega_0^2(1+A\cos(\omega_m t))x=0$ where $x$ is the displacement, $\omega_m$ and $A$ are the frequency and amplitude of the external driving and $\omega_0$ is a natural frequency of the system~\cite{Kovacic}. Floquet analysis of the Mathieu equation gives a series of resonances that occur at $\omega_m=2\omega_0/n$ where $n$ is an integer~\cite{Barone}.

Bose-Einstein condensates (BECs) of atomic gases offer a versatile nonlinear quantum system for the exploration of collective modes and non-equilibrium dynamics due to their unprecedented controllability. The equation governing the dynamics of a weakly interacting condensate at zero temperature is the well-known Gross-Pitaevskii (GP) equation \cite{pitaevskii-16}, which, in appropriate limits, takes the form of a Mathieu-like equation, allowing parametric amplification of collective waves. During the last two decades,  pattern formation in BECs has been explored both theoretically and experimentally by periodically modulating the confining potential~\cite{Dalfovo1,Dalfovo2,Dalfovo3,Nicolin,Engels,Nicolin2,Jaskula,Smits} or the interaction parameters~\cite{Brazhnyi,Kartashov,Staliunas,Vidanovic,Okazaki,Nguyen,Kwon,Maity,Zhang}. For example, Engels et al. observed one-dimensional (1D) Faraday waves that exhibit a longitudinal density oscillation by periodically modulating the transverse confining potential in a single-component BEC~\cite{Engels}. Nguyen et al. observed the transition of Faraday patterns to granulation in an elongated single-component BEC by periodically modulating the interaction~\cite{Nguyen}. Kwon et al. observed star-shaped surface patterns in a circular single-component BEC by periodically modulating the scattering length near the Feshbach resonance~\cite{Kwon}. Zhang et al. observed density patterns with two-, four-, and six-fold symmetries when the atomic interactions were modulated at multiple frequencies~\cite{Zhang}. Faraday waves have been recently observed also in Fermi superfluids~\cite{Rajkov}.    

The realization of mixtures of Bose gases occupying different hyperfine states or atomic species opens up the possibility of exploration of spin-related phenomena~\cite{Recati,Baroni}. In a binary BEC, the trapping potentials of the two components can be different. Depending on the relative value of the inter- and intra-spin interaction constants, the two components can be in a miscible or immiscible phase~\cite{Recati,Baroni, Papp, Wacker}. Furthermore, the coherent coupling between the two components can be produced via Radio Frequency or Raman transitions, giving rise to Rabi-coupled~\cite{Matthews} and spin-orbit-coupled BEC, allowing for the exploration of interesting phenomena like, for instance, topology physics~\cite{Lin} or spin domains spontaneously generated in a quench through the miscible-immiscible phase transition~\cite{Sabbatini}. 

Recently, Cominotti et al. observed the formation of total-density and spin-density Faraday waves by modulating the transverse confinement potential for an elongated two-component BEC~\cite{Cominotti}. This experiment was carried out with sodium atoms prepared in the 
$|F=1; m_F=\pm 1 \rangle$ internal states. The fact that the condensate is near the miscible-immiscible phase transition allows for the separation of the total density and spin density excitations. Under the same condition, many spin-related effects, such as magnetic solitons~\cite{Chai,Farolfi,Qu}, spin-dipole polarizability and spin-dipole mode~\cite{Bienaimé}, and spin sound~\cite{Kim}, have been investigated in the last few years.

Motivated by the recent experiment of Ref.~\cite{Cominotti}, we numerically investigate parametric excitations in a binary BEC near the miscible-immiscible phase transition. For an elongated BEC, in addition to the generation of total density and spin density Faraday patterns, we explore the fate of spin Faraday patterns when the system enters a nonlinear regime at a longer driving time. For a pancake-shaped BEC, we obtain spin Faraday patterns with an $l$-fold rotational symmetry, where the number of nodes along the angular and radial directions of the pattern depends on the modulation amplitude and modulation frequency. 

This paper is structured as follows. In Sec.~\ref{sec-model}, we introduce the coupled GP equations to simulate the dynamics of a binary BEC and describe the modulation schemes for parametric excitations. In Sec.~\ref{sec-1D}, we demonstrate the formation of spin Faraday patterns in an elongated BEC and the appearance of a nonlinear destabilization regime. Section~\ref{sec-2D} is devoted to the generation of two-dimensional (2D) spin Faraday patterns in a pancake-shaped BEC. Section~\ref{sec-conclusion} is our conclusion.

\section{System model}
\label{sec-model}

We consider a two-component BEC at zero temperature,  confined in a harmonic trap of frequencies $\omega_x, \omega_y$, and $\omega_z$.  For example, the two components can consist of the $|F=1; m_F=\pm 1 \rangle$ hyperfine states of $^{23}$Na atoms. In these conditions, the condensate is governed by the following three-dimensional (3D) GP equations  \cite{pitaevskii-16}
\begin{eqnarray}
i \hbar \frac{\partial \psi_1}{\partial t} &=& \left( 
-\frac{\hbar^2 \nabla^{2}}{2m} +V+\ g_{11}|\psi_{1}|^{2} 
+g_{12}|\psi_{2}|^{2} \right) \psi_{1} \ , \ \ 
\label{eq:GPE1} \\
i \hbar \frac{\partial \psi_2}{\partial t} &=& \left( 
-\frac{\hbar^2\nabla^{2}}{2m}+V+ g_{22}|\psi_{2}|^{2} 
+g_{12}|\psi_{1}|^{2} \right) \psi_{2} \ ,  
\label{eq:GPE2} 
\end{eqnarray}
where $\psi_{j=1,2}$ denotes the order parameter of each component, satisfying the normalization condition $\int dx dy dz\ n_{j}(x,y,z) = N_j$, where $n_j=|\psi_{j}(x,y,z)|^{2}$ is the density and $N_j$ is the atom number. We assume that the two components are equally populated, so that $N_j=N/2$. For the harmonic confining potential, $V= \frac{1}{2}m(\omega_{x}^{2}x^{2}+\omega_{y}^{2}y^{2}+\omega_{z}^{2}z^{2})$, we consider two different geometries: (i) an elongated BEC with $(\omega_x, \omega_y,\omega_z)=2\pi\times\{5, 512, 512\}$~Hz, and (ii) a pancake-shaped BEC with $(\omega_x,\omega_y,\omega_z)=2\pi\times\{50, 50, 1500\}$~Hz.

The problem can be mapped into a spin-$1/2$ system, where $n_1$ and $n_2$ are the densities of spin-up and spin-down particles, respectively. The total density can be defined as $n=n_1+n_2$ and the spin density as $n_s=n_1-n_2$, and $n_s/n$ has the meaning of local magnetization of the system.  
 The effective 3D interaction constants $g_{ij}$ can be determined from the $s$-wave scattering lengths $a_{ij}$ using the relation $g_{ij} = 4\pi\hbar^2 a_{ij}/m$. For the two selected internal states of $^{23}$Na atoms, the intra-component interaction constants are the same, $g_{11}=g_{22}\equiv g$, and the inter-component interaction constant is slightly smaller, $g_{12}=0.93 g$. Thus, the system is in the miscible regime, $g_{12}^2<g_{11}g_{22}$, but close enough to the miscible-immiscible phase transition where the total density and spin density excitations are decoupled~\cite{Pu,Timmermans}. Moreover, the system is dynamically stable, that is, the frequencies of all Bogoliubov excitations obtained by linearizing the GP equations (\ref{eq:GPE1})-(\ref{eq:GPE2}) are real.

As already mentioned in the Introduction, parametric amplification can be obtained by periodically modulating either the harmonic trapping potential or the scattering lengths. In this work, we consider a modulation of the intra-component scattering lengths in the form 
\begin{eqnarray}
a_{11}(t) & = & a + a_{m} \sin(\omega_{m} t)  \nonumber \\ 
a_{22}(t) & = & a \pm a_{m} \sin(\omega_{m} t)\ ,
\label{a_B}
\end{eqnarray} 
where $a_{m}$ and $\omega_m$ are the modulation amplitude and frequency, while $a_{12}$ remains constant. The case with positive sign in the second expression corresponds to an \textit{in-phase} modulation of the scattering lengths, while the case with negative sign corresponds to an \textit{out-of-phase} modulation. Both produce parametric amplification, but the effectiveness with which they do so can be different. In experiments, modulation of the scattering lengths can be obtained via Feshbach resonances by applying an oscillating magnetic field to the atoms. By modulating the trapping frequency at fixed interaction strengths, very similar effects would be obtained. In fact, modulating the trapping frequency induces a periodic variation in density; but the nonlinear term in the GP equation that causes parametric amplification contains the product of density and interaction strength. Therefore, within GP theory, the choice of modulating one factor or the other is a matter of convenience, since the effects are the same. 

Patterns form when parametrically amplified collective excitations produce visible standing waves, whose shape depends on the resonance conditions and geometry of the system. To be amplified, these collective excitations must already be present in the initial configuration in the form of initial noise. In our simulations, the presence of random numerical noise is guaranteed by the fact that the order parameters of the two components at initial time are the stationary solutions of GP equations (\ref{eq:GPE1})-(\ref{eq:GPE2}), which we calculate using an imaginary time iterative method. These solutions are never exact, because the iterations stop at a certain point, close to convergence but not exactly at it. The remaining small deviations from the exact ground state are sufficient to trigger parametric resonance in the subsequent real-time evolution.  As we will do in Section IV, we can also manually add extra random noise to the initial order parameters, when useful. The amount of initial noise affects the time scale at which the pattern becomes visible, but not its shape and resonance condition (i.e., the relationship between the modulation frequency and the amplified excitation frequency). In experiments, there is always some level of noise that can facilitate the appearance of a pattern in a relatively short time. 

\section{Faraday pattern in an elongated BEC}
\label{sec-1D}

If $\omega_{y,z} \gg \omega_x$, the condensate is very elongated along $x$.  For the total number of atoms we choose $N=10^5$. With this value, the total density $n$ turns out to be well approximated by the inverted parabola obtained with the Thomas-Fermi approximation~\cite{pitaevskii-16}, $n_{\rm TF}(x,y,z)=(\mu - V(x,y,z))/g$, where $\mu$ is the chemical potential. At frequencies lower than $\omega_{x,y}$, the dynamical response to external perturbations is dominated by collective axial waves. Instead of numerically calculating the full spectrum of Bogoliubov excitations by linearizing the coupled GP equations (\ref{eq:GPE1})-(\ref{eq:GPE2}), it is convenient to approximate the lowest-energy dispersion law of axial phonons  by using the gapless branches
\begin{equation}
\omega_{d,s} = \sqrt{\frac{k_x^2}{2m}\left(\frac{ \hbar^2 k_x^2}{2m} + \bar{n}(g \pm g_{12}) \right)} \ ,
\label{eq:Bog}
\end{equation}
for the density and spin channels, respectively. This expression is the usual Bogoliubov phonon dispersion in a uniform condensate~\cite{Pethick}, except for the use of an average density $\bar{n}$ accounting for the inhomogeneity of the trapped condensate. The inhomogeneity along the $x$ direction gives small effects on the dispersion, whereas the role of the tight transverse confinement in the $yz$-plane has been investigated in~\cite{E.Taylor, Kramer} for infinite cylindrically-symmetric condensates. In the Thomas-Fermi regime, it was found that  $\bar{n}=n(0,0,0)/2$, where $n(0,0,0)$ is the central density. We will use this value to obtain an analytic estimate of the resonance condition for parametric amplification.

In the long wavelength limit, the dispersion reduces to
\begin{equation}
    \omega_{d,s} = c_{d,s}k_x \ ,
    \label{eq:lineardisp}
\end{equation}
where $c_{d,s}=\sqrt{\bar{n}(g\pm g_{12})/2m}$ corresponds to the velocity of sound and spin sound, respectively. Near the miscible-immiscible phase transition, the spin sound velocity $c_s$ is much smaller than the density sound velocity $c_d$ while the spin healing length $\xi_s=\hbar/\sqrt{2m \bar{n} (g-g_{12})}$ is much larger than the density healing length $\xi_d=\hbar/\sqrt{2m \bar{n} (g+g_{12})}$. The two distinct sound modes in the density and spin channels enable the generation of density Faraday patterns and spin Faraday patterns separately.

\subsection{Total density Faraday pattern}

First, we discuss total density Faraday patterns generated by an \textit{in-phase} modulation of the scattering lengths, with $a_{11}(t)=a_{22}(t)=a+a_m\sin(\omega_m t)$ and $a_{12}=0.93a$.  At $t=0$, the condensate wave function is the stationary solution of Eqs.~(\ref{eq:GPE1})-(\ref{eq:GPE2}). 

\begin{figure}[t!]
\includegraphics[width=0.49\textwidth]{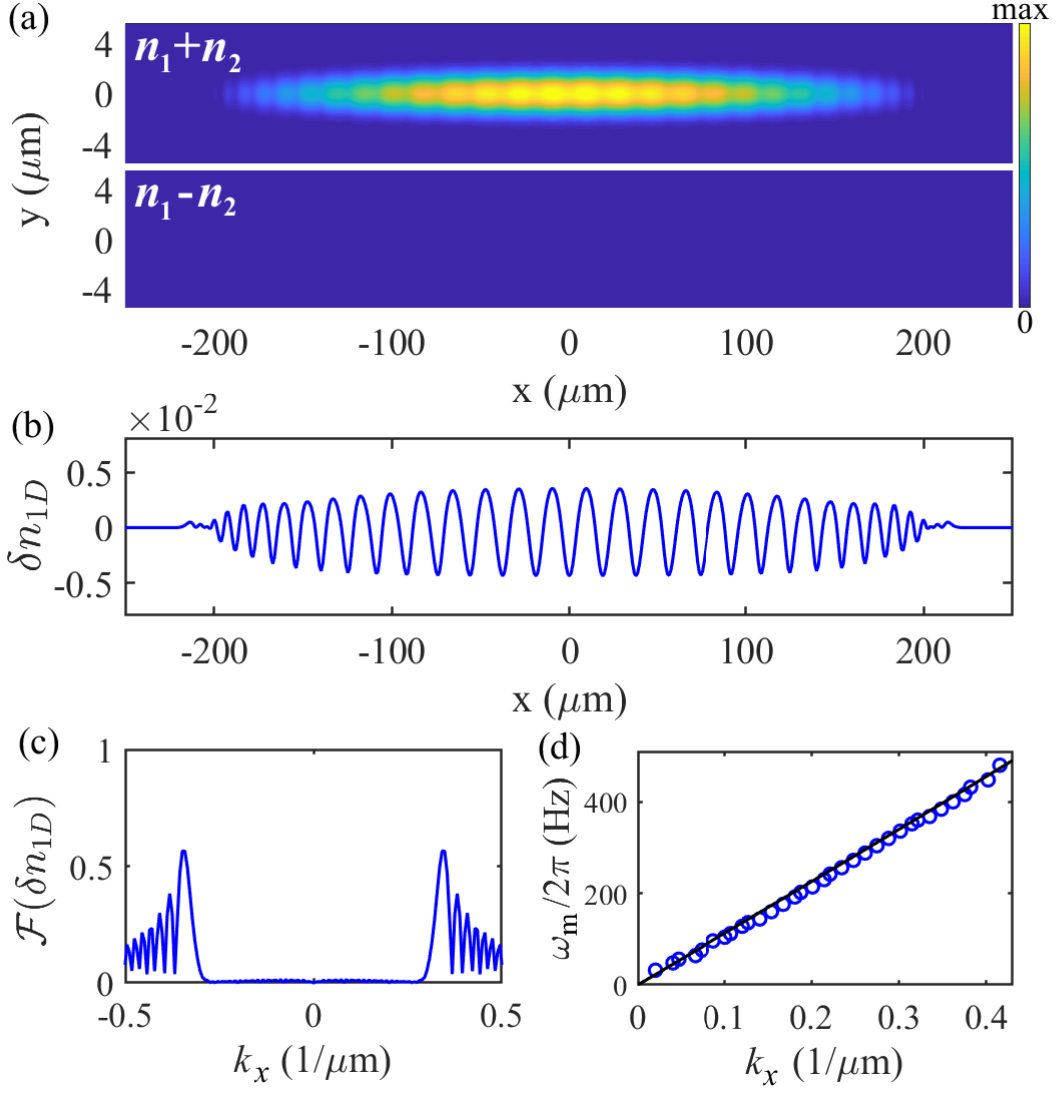} 
\centering
\caption{Total density Faraday pattern in an elongated two-component BEC generated by an \textit{in-phase} modulation of the scattering lengths $a_{11}$ and $a_{22}$, starting from the ground state at $t=0$. (a) Total density $n=n_1+n_2$ and spin density $n_s= n_1-n_2$, in the $xy$-plane at $t=495$ ms. (b) Integrated 1D density $\delta n_{1D}(x)=\iint dy dz\ (n-n_{\rm TF})$, after subtracting the Thomas-Fermi equilibrium density. (c) Fourier transform of $\delta n_{1D}(x)$; the side peaks at $k_x\simeq \pm 0.35$ $\mu$m$^{-1}$ are associated with the counter-propagating Bogoliubov modes producing the standing density wave. For panels (a-c), the modulation frequency and amplitude are $\omega_{m}/2\pi=384$~Hz and $a_{m}=0.36a$, respectively. (d) Relation between $\omega_m$ and the corresponding side peak wave vector $|k_x|$; the blue circles correspond to the results extracted from the numerical simulations; the solid line is the resonant condition for parametric excitation,  $\omega_m=2\omega_d(k_x)$, where $\omega_d(k_x)$ is the dispersion law Eq.~(\ref{eq:Bog}), which is indistiguishable from the linear dispersion $2c_d k_x$ in this range of $k_x$. }
\label{FIG.1}
\end{figure}

Figure~\ref{FIG.1}(a) shows the typical distributions of the total density $n$ and the spin density $n_s$ after applying a modulation of frequency $\omega_{m}=2\pi\times384$~Hz and amplitude $a_{m}=0.36a$, for a duration time $t\simeq 495$~ms. A clear Faraday pattern is visible in the total density, while the spin density remains unaffected. In Fig.~\ref{FIG.1}(b), we show the integrated 1D density profile $\delta n_{1D}(x)=\iint dy dz\ (n-n_{\rm TF})$ after subtracting the equilibrium Thomas-Fermi density; its spatial oscillation corresponds to a standing wave of counter-propagating Bogoliubov modes in the density channel. 

\begin{figure}[t!]
\includegraphics[width=0.5\textwidth]{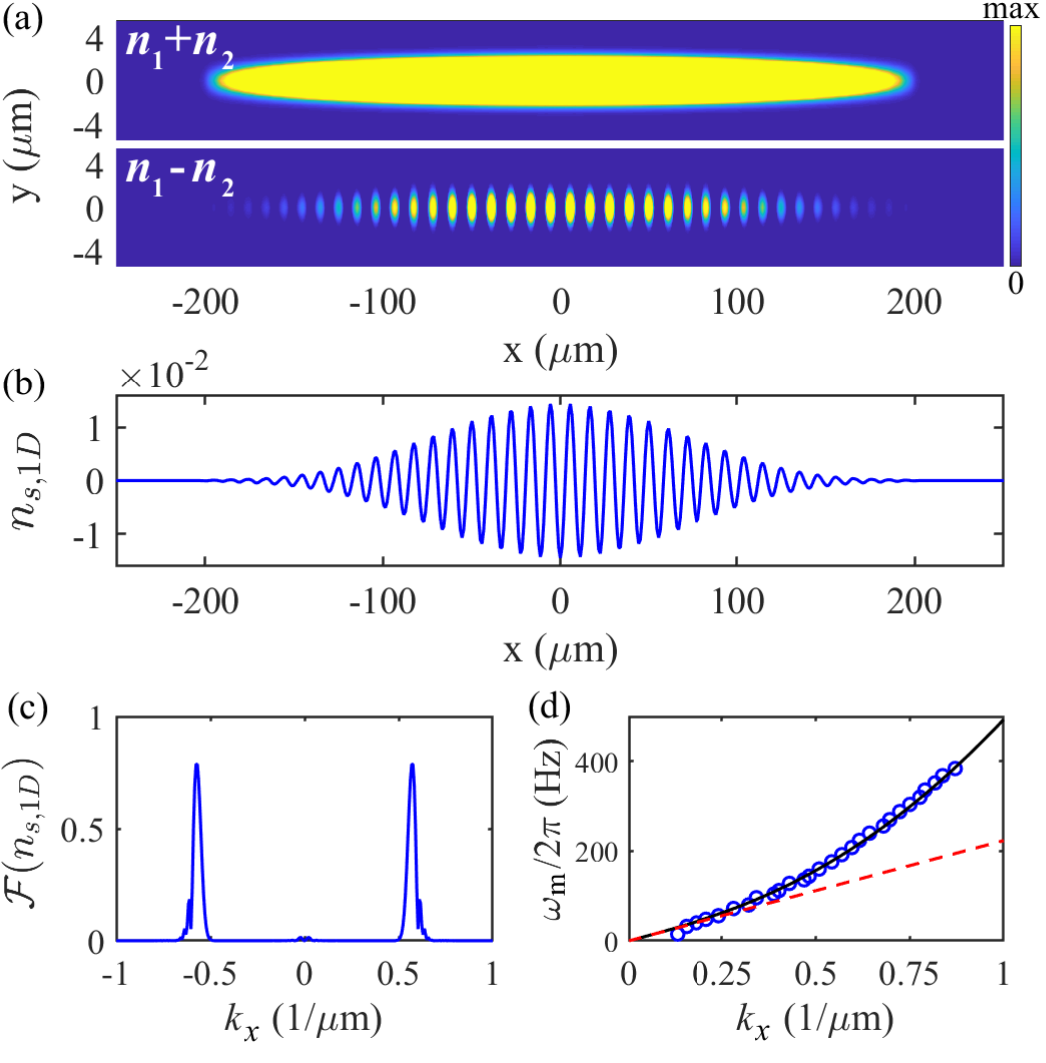} \centering
\caption{Spin density Faraday pattern in an elongated two-component BEC generated by an \textit{out-of-phase} modulation of the scattering lengths $a_{11}$ and $a_{22}$. (a) Total density $n=n_1+n_2$ and spin density $n_s=n_1-n_2$, in the $xy$-plane at $t=219$~ms. (b) Integrated 1D spin density $n_{s,1D}=\iint dy dz\ n_s$. (c) Fourier transform of the $n_{s,1D}$; the side peaks at $k_x\simeq \pm 0.58$ $\mu$m$^{-1}$ are associated with the counter-propagating spin waves producing the pattern. For panels (a-c), the modulation frequency and amplitude are $\omega_m/2\pi=195$ Hz and $a_m=0.07a$, respectively. (d) Relation between $\omega_m$ and the corresponding side peak wave vector $|k_x|$; the blue circles correspond to the results extracted from the numerical simulations; the solid line is the resonant condition for parametric excitation,  $\omega_m=2\omega_s(k_x)$, where $\omega_s(k_x)$ is the Bogoliubov dispersion Eq.~(\ref{eq:Bog}) for the spin branch; the dashed line, is the linear dispersion $2c_s k_x$, valid in the long wavelength limit.}
\label{FIG.2}
\end{figure}

To quantify the wavelength of the pattern, we perform a Fourier transform of $\delta n_{1D}(x)$. As shown in Fig.~\ref{FIG.1}(c), we find two side peaks with maxima at $k_x\simeq \pm 0.35$ $\mu$m$^{-1}$, corresponding to the wavelength $\lambda=2\pi/|k_x|=17.95$~$\mu$m. Repeating the same procedure for different modulation frequencies $\omega_m$, we obtain the relation between the modulation frequency and the wave vector of the peaks. The results are reported as blue circles in Fig.~\ref{FIG.1}(d). These can be compared with the prediction $\omega_m=2\omega_d(k_x)$ for the modulation frequency at which Bogoliubov modes are parametrically excited in the density channel; using Eq.~(\ref{eq:Bog}), this condition is represented by the solid line. The agreement is very good. 

It is worth adding a few comments on Fig.~\ref{FIG.1}. First, note that we limit the simulations to frequencies below the transverse trapping frequency, in order not to excite modes in the $yz$ plane. Second, we remark that the Bogoliubov dispersion is indistinguishable from its linear approximation $2c_d k_x$ in the frequency range considered here; in fact, deviations from linearity are expected when $k_x$ becomes of the order of the inverse of the healing length, and, with our parameters, the healing length is $\xi_d=0.38 \ \mu$m, so $k_x \ll 1/\xi_d$ in the entire range of Fig.~\ref{FIG.1}(d). Third, the two lateral peaks in Fig.~\ref{FIG.1}(c) are broad and asymmetric. This is mainly due to the effect of harmonic trapping in the $x$ direction. Since the system is finite and inhomogeneous, the wave vector $k_x$ is not a good quantum number. The dispersion law (\ref{eq:Bog}) is an excellent approximation for the frequency of the lowest branch in the Bogoliubov excitation spectrum, but it does not take into account density oscillations in the outer part of the condensate, where the density is low. The oscillations in Fig.~\ref{FIG.1}(b) clearly extend into low-density regions far from the center, and this causes the appearance of lateral components of the Fourier transform $\mathcal{F}(\delta n_{1D})$, with $|k_x|$ greater than the wave vector of the phonons propagating in the central region of the condensate, for which Eq.~(\ref{eq:Bog}) applies. This interpretation is also consistent with the time dependence of the peak structure. We observe that parametric amplification begins in the central region and that two isolated narrow peaks appear where the resonance condition given by Eq.~(\ref{eq:Bog}) is fulfilled; only when the entire condensate is excited, the peaks become broader and more asymmetric. At a later time, nonlinear mode coupling causes further fragmentation of $\mathcal{F}(\delta n_{1D})$, as we will discuss later also for spin excitations.

\subsection{Spin density Faraday pattern}

We generate spin density Faraday patterns by applying an \textit{out-of-phase} modulation of the intra-component scattering lengths in the form $a_{11}(t)=a+ a_{m}\sin(\omega_{m} t)$ and $a_{22}(t)=a - a_{m}\sin(\omega_{m} t)$. We use a modulation frequency $\omega_{m}=2\pi\times 195$~Hz, while the  amplitude is chosen to be $a_{m}=0.07a$, such that miscible condition for the ground state ($g_{12}^2 < g_{11}g_{22}$) remains valid within the whole interval of scattering lengths.  Figure~\ref{FIG.2}(a) shows the typical distribution of the total density $n$ and spin density $n_s$ after applying the modulation for a time $ t=219$~ms. In contrast to the total density Faraday pattern shown in Fig.~\ref{FIG.1}, we find that the total density is barely excited while the spin density exhibits a regular pattern along the longitudinal direction. 

In Fig.~\ref{FIG.2}(b) we show the integrated 1D spin density profile $n_{s,1D} = \iint dy dz \ n_s$. The spatial oscillation indicates the formation of a standing wave caused by counter-propagating Bogoliubov spin modes. Its Fourier transform, reported in Fig.~\ref{FIG.2}(c), exhibits two side peaks at $k_x\simeq \pm 0.58$ $\mu$m$^{-1}$ corresponding to the wavelength $\lambda=2\pi/|k_x|=10.8\ \mu$m. As already done in the density channel, we can compare the relation between $\omega_m$ and $k_x$ with the resonant condition $\omega_m=2\omega_s(k_x)$ for parametric excitation of spin Bogoliubov modes, where $\omega_s(k_x)$ is the Bogoliubov dispersion Eq.~(\ref{eq:Bog}) in the spin channel. The agreement is again very good. Here, the linear dispersion Eq.~(\ref{eq:lineardisp}) is valid within a narrower frequency range, in line with the fact that the spin healing length is $\xi_s=1.97 \ \mu$m, and is larger than the density healing length $\xi_d$. 

It is worth noting that near the miscible-immiscible phase transition, the spin sound velocity is significantly smaller than the density sound velocity. Consequently, to generate a spin Faraday pattern of a similar wave vector, the required modulation frequency is smaller than that for the generation of a pattern in the total density. Our results are consistent with the experimental observations reported in~\cite{Cominotti}, where the trapping potential was modulated instead of the scattering length. 

\begin{figure}[t!]
\includegraphics[width=0.5\textwidth]{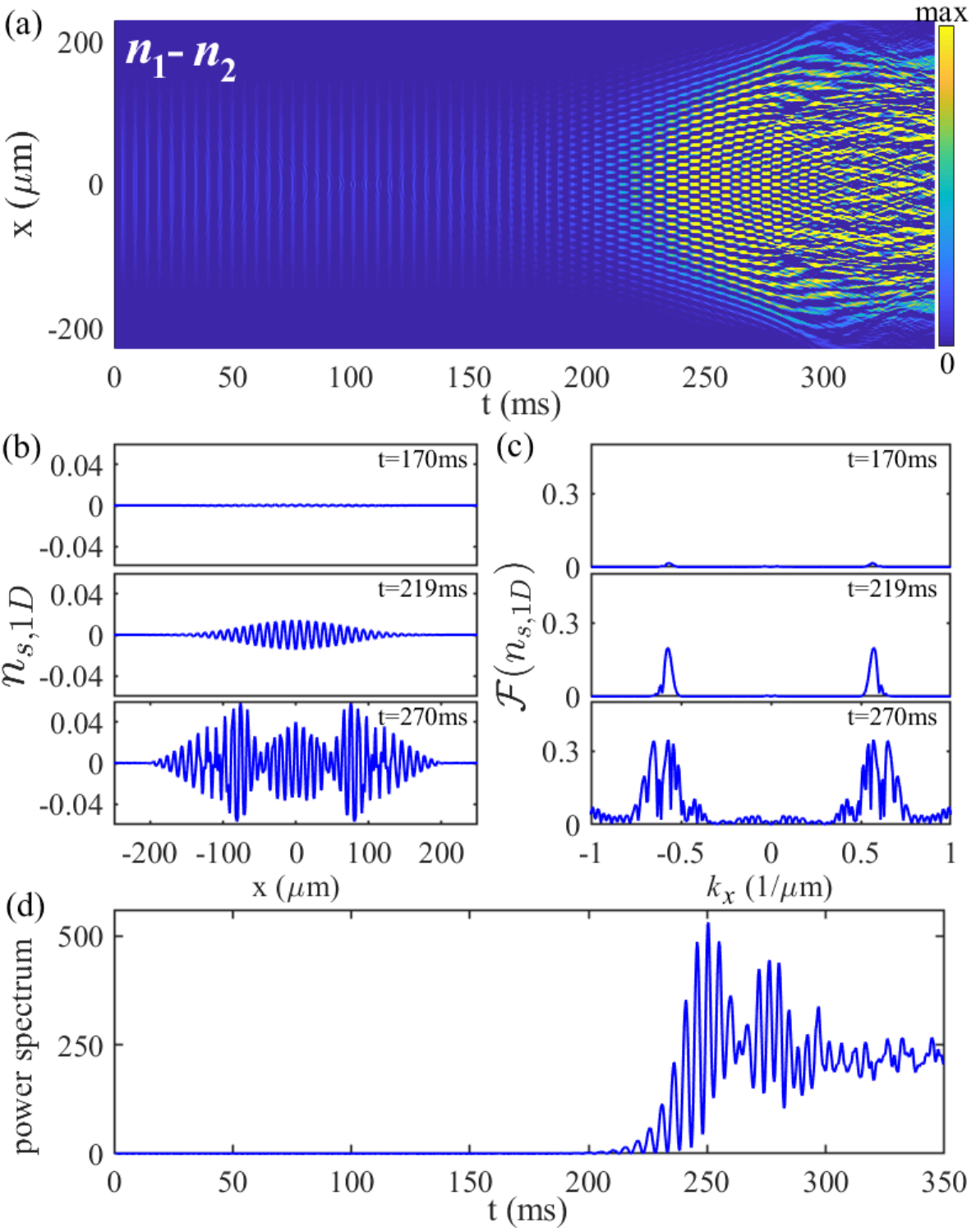}
\centering
\caption{Time evolution of an elongated two-component BEC subject to an \textit{out-of-phase} modulation of the scattering lengths $a_{11}$ and $a_{22}$, started at $t=0$, with modulation frequency $\omega_{m}=2\pi\times 195$~Hz and  amplitude $a_{m}=0.07a$. (a) Integrated 1D spin density $n_{s,1D} (x)$. (b-c) Integrated 1D spin density profiles at $t=170, 219, 270$~ms, and the corresponding Fourier transforms. (d) Time evolution of the integrated power spectrum, $ \int_{-\infty }^{+\infty } dk_x |\mathcal{F}(n_{s,1D})|^2$, of the spin density profile. }
\label{FIG.3}
\end{figure}

In addition to its generation, we also explore the fate of the spin Faraday pattern after a long dynamic evolution under periodic modulation. In Fig.~\ref{FIG.3}(a) we show the complete temporal evolution of the integrated 1D spin density, with the same parameters as in Fig.~\ref{FIG.2}. The condensate shows tiny oscillations for about 200 ms from the start of modulation, then the Faraday pattern of the spin density appears and becomes increasingly visible. The time required for the pattern to become visible depends on the initial amount of noise. Once visible, its amplitude increases rapidly. Its evolution can be better interpreted by observing panels (c)-(d) of the same figure. In particular, Fig.~\ref{FIG.3}(d) shows the integrated power spectrum, $\int_{-\infty }^{+\infty } dk_x |\mathcal{F}(n_{s,1D})|^2$, where $\mathcal{F}(n_{s,1D})$ is the Fourier transform of the spin density. This quantity shows exponential growth when the Faraday pattern forms, but then saturates. This saturation is accompanied by a randomization of the density distribution, which begins in the outer parts of the condensate and gradually spreads throughout the volume. The time interval between the initial appearance of the Faraday pattern and the moment when the power spectrum is maximum is about $50$~ms and is approximately equal to the time required for the spin waves to propagate from the center to the boundary. We have verified numerically that this time interval increases when the trapping frequency $\omega_x$ is brought to $3$~Hz instead of $5$ (i.e., for a larger condensate) by a value in line with expectations. Ultimately, the long-term evolution is dominated by nonlinear coupling between many spin modes, both in the bulk and at the boundaries, and parametric amplification is no longer effective.    

\begin{figure}[t!]
\includegraphics[width=0.5\textwidth]{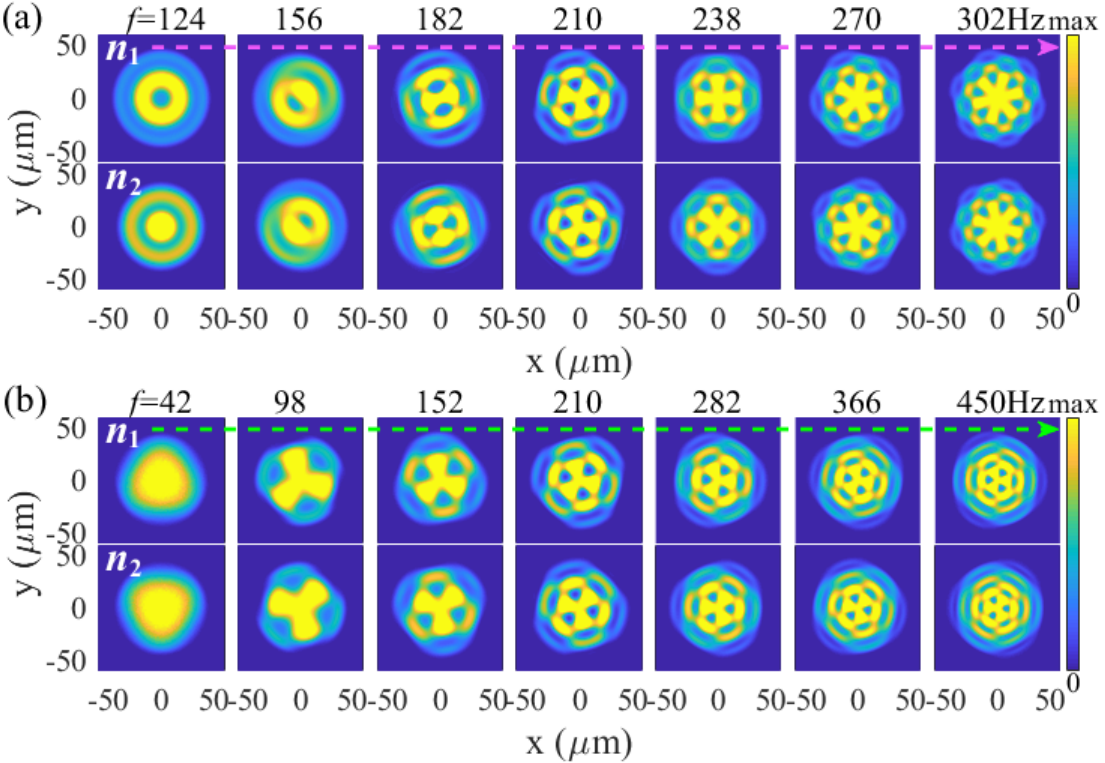} 
\centering
\caption{Spin Faraday patterns in a two-component pancake-shaped BEC with increasing angular $l$ and radial $n_r$ quantum numbers. In panel (a) we plot the density distributions $n_1$ and $n_2$ for patterns with three radial nodes ($n_r=3$) and increasing values of the angular momentum from $l=0$ to $l=6$, from left to right; these patterns are obtained with modulation frequencies $f\equiv \omega_{m}/2\pi=124, 156, 182, 210, 238, 270, 302$~Hz at $t=320, 956, 1316, 951, 898, 877, 897$~ms. The configurations in panel (b) have $l=3$ and increasing radial quantum number from $n_r=0$ to $n_r=6$, with modulation frequencies $f\equiv\omega_{m}/2\pi=42, 98, 152, 210, 282, 366, 450$ Hz at $t=2895, 1377, 1032, 947, 829, 748, 806$~ms. The modulation amplitude is $a_{m}=0.037a$, except in the first column of the panel (b), where it is $a_{m}=0.07a$. }
\label{FIG.4}
\end{figure}

\section{Spin Faraday pattern in a pancake-shaped BEC}
\label{sec-2D}

In this section, we investigate the formation and evolution of spin Faraday patterns in a pancake-shaped BEC. Similarly to the case of the elongated BEC, we apply \textit{out-of-phase} periodic modulations to the scattering lengths, starting from the stationary state at $t=0$. We also use \textit{in-phase} modulations, and we add a certain amount of noise to the initial condensate wave function. As we will discuss later, all protocols lead to the formation of Faraday patterns, though with different efficiency. 

The onset of parametric amplification in a pancake-shaped BEC is indicated by the formation of 2D patterns that exhibit an $l$-fold rotational symmetry with $n_r$ nodes along the radial direction. As shown in Fig.~\ref{FIG.4}(a) and (b), as the modulation frequency increases, spin Faraday patterns with increasing radial and angular quantum numbers $(n_r, l)$ emerge when proper modulation amplitudes are applied. Specifically, each column of Fig.~\ref{FIG.4}(a) shows the density distributions of the two components for patterns with three radial nodes ($n_r=3$) and increasing values of $l$, corresponding to increasing values of the modulation frequency. Instead, Fig.~\ref{FIG.4}(b) shows the density distributions for a fixed angular momentum $l$, but increasing values of $n_r$. For any given $l$, the system is invariant for a rotation of an angle $2\pi/l$. In all cases, the two densities $n_1$ and $n_2$ are out-of-phase, as expected for a 2D spin Faraday pattern. 
 
\begin{figure}[t!]
\includegraphics[width=0.5\textwidth]{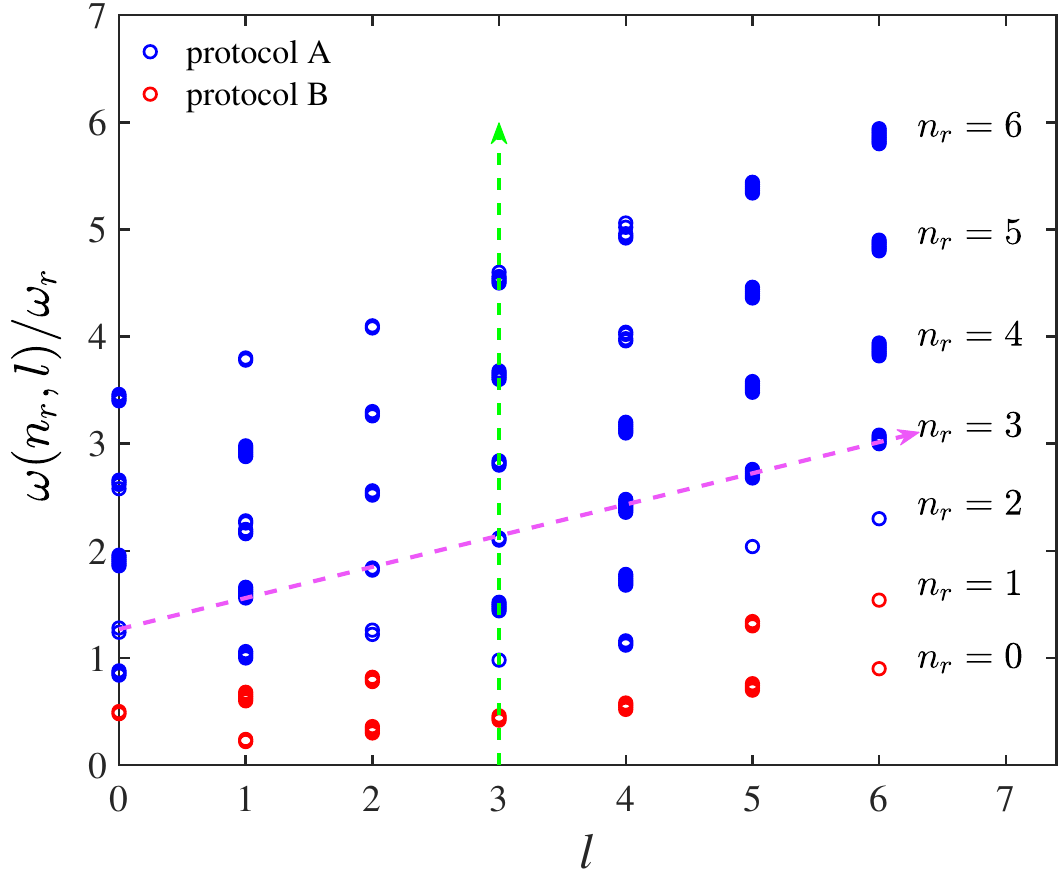}\centering
\caption{Spectrum of spin density excitations, $\omega(n_r,l)$, obtained {\it via} the parametric resonance condition $\omega(n_r,l)=\omega_m/2 $, where $\omega_m$ is the optimal modulation frequency that produces the corresponding Faraday pattern. Frequency is expressed in units of the radial trapping frequency $\omega_r\equiv \omega_x=\omega_y$. Blue circles are obtained with an \textit{out-of-phase} modulation (protocol A), while red circles are with an \textit{in-phase} modulation with initial noise (protocol B).  The points along the oblique and vertical dashed lines correspond to the spin density distributions plotted in Fig.~\ref{FIG.4}(a) and (b), respectively. }
\label{FIG.5}
\end{figure} 
 
To map out the spectrum of parametric resonances, we have performed extensive numerical simulations by solving the coupled GP equations. Among all simulations we select those producing a well-defined spin pattern characterized by the quantum numbers $(n_r,l)$, and we associate each pattern with the optimal modulation frequency $\omega_m$, that is, the frequency at which the resonance occurs. From these optimal frequencies, we can extract the spectrum of spin excitations $\omega(n_r,l)$ by assuming the parametric resonance condition $\omega_m = 2 \omega(n_r,l)$. The spectrum obtained in this way is shown by the circles in Fig.~\ref{FIG.5}. The points along the oblique and vertical dashed lines correspond to the density distributions plotted in Fig.~\ref{FIG.4}(a) and (b), respectively.  
These discrete collective excitations could also be calculated by solving the Bogoliubov equations for the trapped condensate, as done in  \cite{Ticknor1, Ticknor2} for a similar geometry, but a different set of parameters. For a miscible two-component BEC, the excited modes are collective in nature and the two species can move either in-phase (total density waves) or out-of-phase (spin density waves) with each other. For a given pair of quantum numbers $\omega(n_r,l)$, excitation modes in the spin channel generally have lower energy than those in the density channel~\cite{Ticknor2}. 

\begin{figure}[t]
\includegraphics[width=0.5\textwidth]{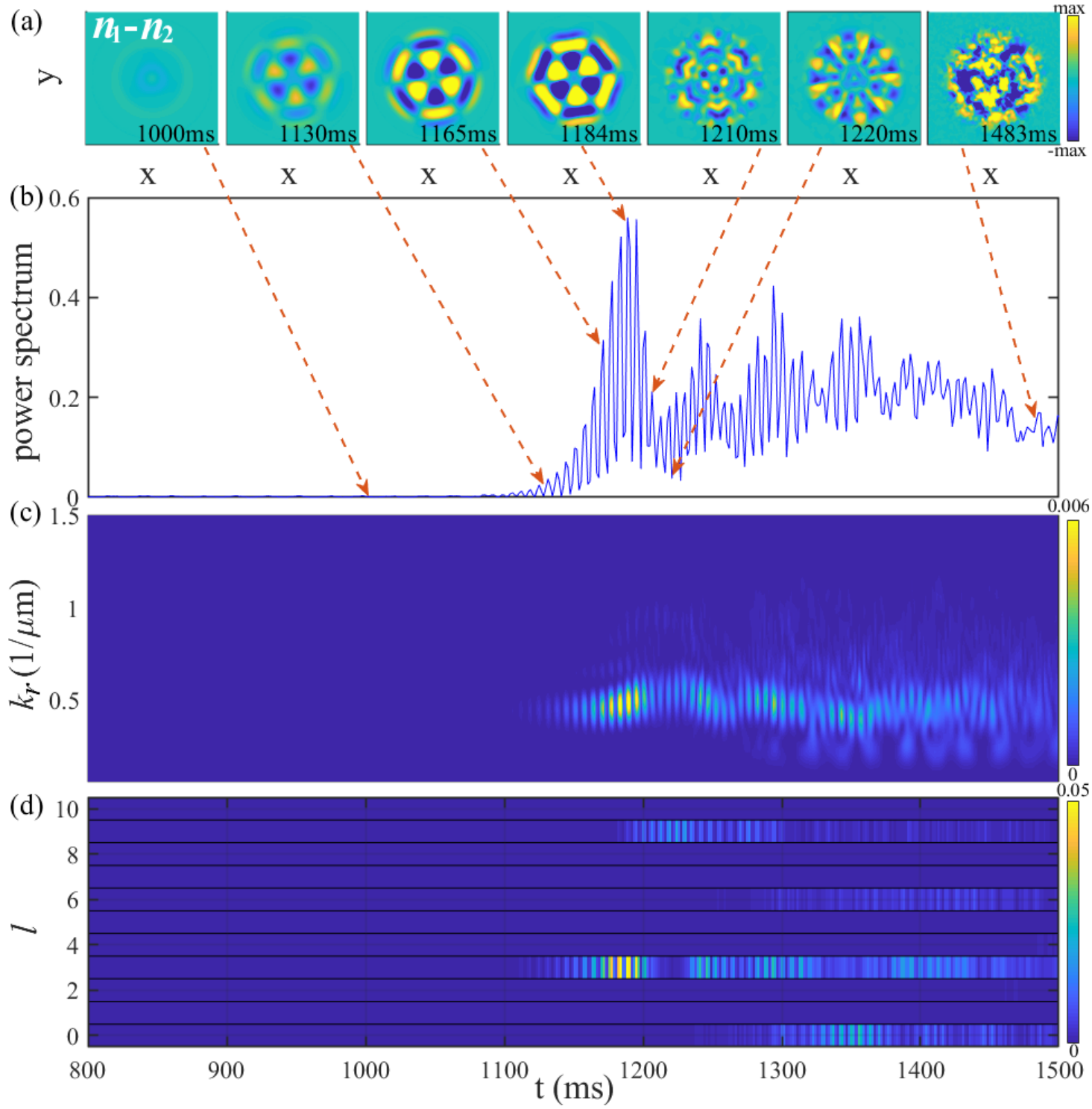}
\centering
\caption{Example of parametric amplification caused by a \textit{out-of-phase} modulation of amplitude $a_m=0.037a$ and frequency $\omega_m/2\pi=152$~Hz. (a) Selected snapshots of the spin density, where one can see the formation of a spin pattern with $l=3$ and $n_r=2$ and its subsequent destruction; each square corresponds to $60 \times 60 \ \mu$m. (b) Time evolution of the power spectrum, $\sum_{l=0}^\infty \int_0^\infty dk_r |P_l(k_r)|^2$, where $P_l(k_r)$ are the coefficients of the decomposition of the 2D spin density on the basis of Bessel functions and eigenfunctions of the angular momentum. (c) Time evolution of the function $\sum_{l} |P_l(k_r)|^2$. (d) Time evolution of the function $\int\! dk_r |P_l(k_r)|^2$. All functions are plotted in arbitrary units, since normalization is irrelevant in this context.}
\label{FIG.6}
\end{figure}

A typical temporal evolution of the spin density during the modulation of the scattering lengths is shown in Fig.~\ref{FIG.6}. The seven plots in panel (a) are selected snapshots of the spin density; they clearly show the appearance and growth of a pattern with $l=3$ and $n_r=2$, and its subsequent disruption. In order to quantify the intensity of the pattern, generalizing what we did in 1D, we perform a Fourier analysis in polar coordinates $(r,\phi)$. In particular, we calculate the quantity $P_l(k_r)= \sqrt{k_r/2\pi} \int_0^{2\pi} d\phi \int_0^{\infty} dr\ r 
n_{s,2D}(r,\phi) J_l (k_r r) e^{-i l\phi}$, where $n_{s,2D}(r,\phi)$ is the spin density integrated in the transverse direction $z$, $J_l$ is the Bessel function of order $l$, and $k_r$ is the radial wave vector. We then define the power spectrum as $\sum_{l=0}^\infty \int_0^\infty dk_r |P_l(k_r)|^2$. The temporal evolution of this quantity is shown in Fig.~\ref{FIG.6}(b), which is the 2D analogue of the 1D results already plotted in Fig.~\ref{FIG.3}(d). In this figure, the sum is performed up to $l=10$ and the integral up to $k_r=1.5\ \mu$m$^{-1}$. Further information can be obtained by analyzing the behavior of the radial and angular contributions separately. In Figs.~\ref{FIG.6}(c) and (d), we report the quantity $\sum_{l} |P_l(k_r)|^2$ as a function of $k_r$, and $\int\! dk_r |P_l(k_r)|^2$ as a function of $l$, respectively, as they evolve in time. Overall, Fig.~\ref{FIG.6} shows that the pattern begins to form slightly after 1100 ms; it grows in intensity for about 100 ms, extending from the center to the edge of the condensate. Then, other modes with different $l$ appear \textit{via} nonlinear mode coupling, starting from $l=9$. Careful analysis of the evolution of the spin density shows that this coupling initially occurs in the outer part of the condensate. The subsequent dynamics gradually involves modes with different values of $l$ and with a wider distribution of wave vectors $k_r$ (corresponding to different numbers of nodes $n_r$). Nonlinear mode coupling is enhanced whenever the frequencies of diﬀerent modes, or of their harmonics, almost coincide~\cite{Minniti}, causing beating effects, collapse-and-revival of the initially excited mode, and, at longer times, a complete randomization of the spin structure.

\begin{figure}[t]
\includegraphics[width=0.4\textwidth]{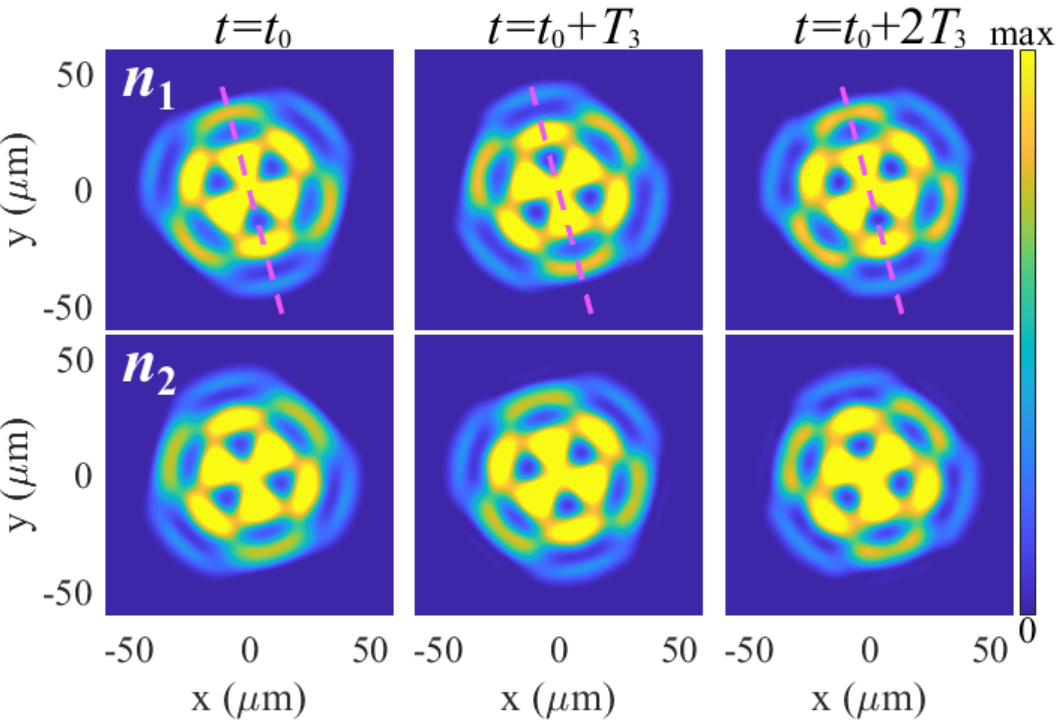}
\centering
\caption{Dynamics of the spin Faraday pattern with $l=3$ and $n_r=3$, generated via modulation with frequency $\omega_m/2\pi =210$ Hz (i.e., modulation period $T_{3}=1/210$ s and modulation amplitude $a_{m}=0.037a$). The pattern oscillates at half of the modulation frequency, indicating the subharmonicity of the spin Faraday pattern.}
\label{FIG.7}
\end{figure}

\begin{figure}[t]
\includegraphics[width=0.5\textwidth]{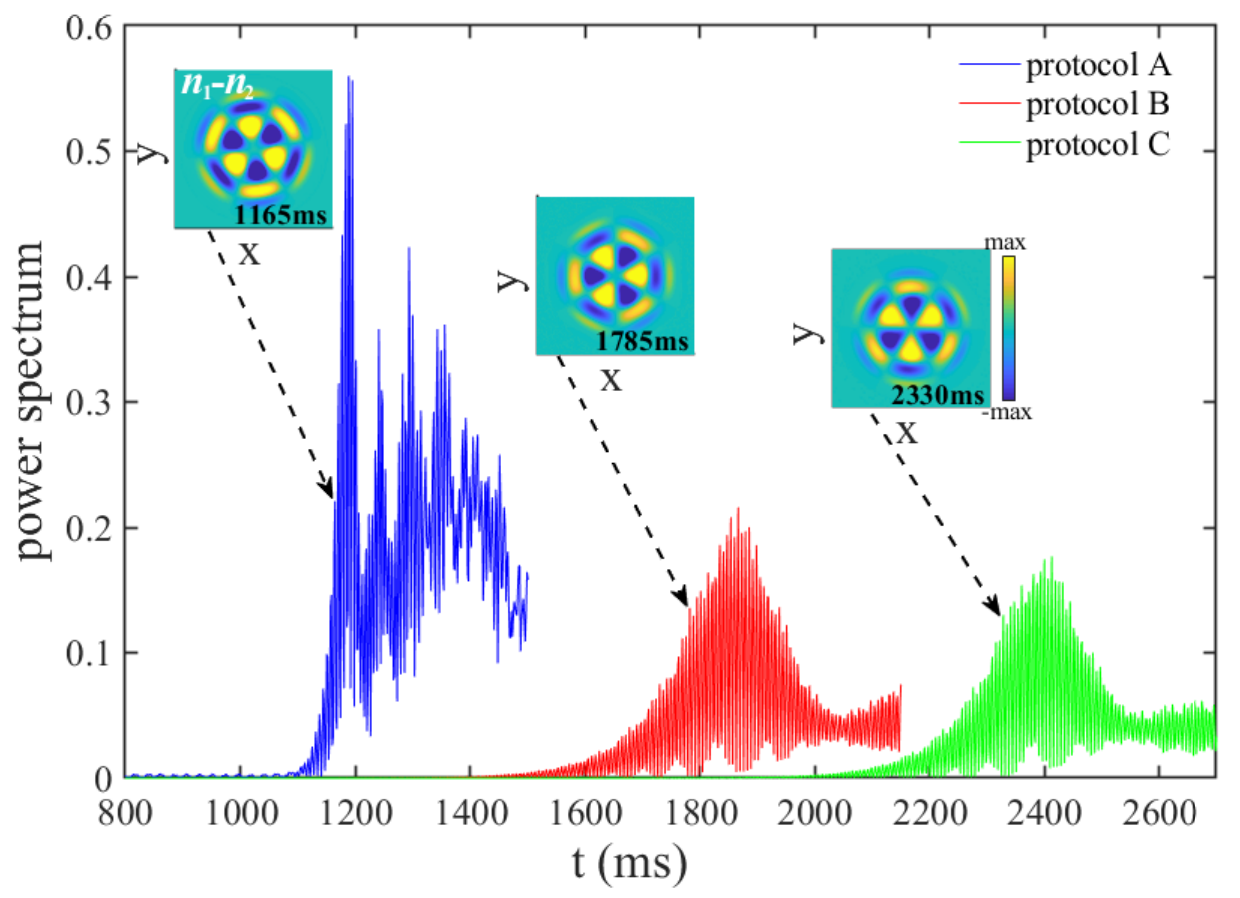}
\centering
\caption{Power spectrum calculated with three different protocols. Protocol A: \textit{out-of-phase} modulation starting from the stationary solution of the GP equations (\ref{eq:GPE1})-(\ref{eq:GPE2}), as in Fig.~\ref{FIG.6}. Protocol B: \textit{in-phase} modulation starting from the stationary solution with added random noise in the modulus of the order parameter. Protocol C: \textit{in-phase} modulation starting from the stationary solution with added random noise in the phase of the order parameter. In all cases, $a_m=0.037a$ and $\omega_m/2\pi=152$~Hz. In both protocols B and C the noise amplitude is $\eta=0.001$. The color plots show the spin density at the instant indicated by the corresponding arrow. }
\label{FIG.8}
\end{figure}

An interesting dynamical feature of the spin Faraday pattern is reported in Fig.~\ref{FIG.7}, where we show the density distributions $n_1$ and $n_2$ of a triangular pattern at some specific time moments $t=t_0$, $t_0+T_3$, and $t_0+2T_3$ where $T_3=2\pi/\omega_m$ is the modulation period. The density configuration rotates by an angle $\pi$ after one modulation period and then by $2\pi$ (i.e., it returns to the original pattern) after two modulation periods, indicating the sub-harmonic nature of the parametric amplification of spin density waves.

As already mentioned at the beginning of this section, the choice of using \textit{in-phase} or \textit{out-of-phase}, with or without additional noise in the initial configuration, is a matter of convenience; in particular, there are no symmetry constraints that prevent the various protocols from working. For all resonances shown as blue circles in Fig.~\ref{FIG.5}, we used \textit{out-of-phase} modulation starting from the stationary solution of the GP equations (\ref{eq:GPE1})-(\ref{eq:GPE2}) (protocol A). We found that modes with $n_r=0$ and $1$ are not easily excitable in this way without adding noise. The reason is easily understood. Spin modes with small $n_r$ and finite $l$ are essentially out-of-phase shape deformations of the spin density or, in other words, surface waves, which leave the central density almost unaffected. In contrast, parametric amplification acts mainly (at least in the initial stages) at the center of the condensate, where the nonlinear terms in the GP equations take on the highest values. Therefore, the coupling between the external modulation and the excitations is weak for these modes, certainly weaker than for high $n_r$. The red circles were obtained with \textit{in-phase} modulation with noise (protocol B). The noise is added in the modulus of the order parameter in the form $\psi _{j} =\sqrt{n_ {0j} }(1+\eta\delta _{j} ) e^{i\phi_{0j} } $, where $n_{0j}$ and $\phi _{0j}$ represent the density and phase of the $j$-th component, $\eta$ is the amplitude of the noise, and $\delta _{j}$ is taken from normally distributed random numbers with a variance of $1$ \cite{Kwon}. We also performed calculations with random noise in the phase (protocol C), such as $\psi _{j} =\sqrt{n_{0j} }e^{i\phi _{0j}(1+\eta\delta _{j} )}$. In real experimental conditions, noise is present in both amplitude and phase. A comparison between the different protocols is shown in Fig.~\ref{FIG.8} for the same Faraday pattern already shown in Fig.~\ref{FIG.6}, with $a_m=0.037a$ and $\omega_m/2\pi=152$~Hz. The blue curve is the same as in Fig.~\ref{FIG.6} with protocol A; the red and green lines are obtained with protocols B and C, respectively, with  noise amplitude $\eta=0.001$. The fundamental characteristic is that the Faraday spin patterns that forms at resonance is the same, with $l=3$ and $n_r=2$, regardless of the protocol. The differences concern the time required for the pattern to appear, its visibility, and its duration. To make the pattern appear earlier, it is sufficient to add even more noise. However, initial noise also modifies the long-term evolution of the spin density, as demonstrated by the different shape of the curves to the right of the maximum of the power spectrum. In the case of protocol A, when the pattern begins to be visible and grows to its maximum visibility, no other modes are present in the condensate, and a few modes gradually come into play through nonlinear coupling, as seen in Fig.~\ref{FIG.6}. In contrast, with protocols B and C, during the formation of the spin pattern, many modes are already present and can exchange energy with the resonant one, leading to a faster decay of the power spectrum. To better understand this long-term dynamics, one could extend the method used recently in \cite{Liebster}, where the stabilization and instability mechanisms of Faraday patterns in a single-component condensate were studied, taking into account the interactions between the excitations generated by the external modulation in the regime of large population of Bogoliubov modes. However, extending this method to the spin and density excitations of a two-component condensate does not seem trivial.

\section{Conclusions}
\label{sec-conclusion}

In this work, we have investigated the parametric excitation and pattern formation in a harmonically trapped binary BECin the miscible regime near the miscible-immiscible phase transition, where the total density and spin density excitations are decoupled. The temporal modulation leads to a symmetry breaking in the spatial density distribution. By periodically modulating the intra-species atomic scattering lengths via \textit{out-of-phase} modulation or \textit{in-phase} modulation with some noise, a spin pattern can be generated with the two components exhibiting an \textit{out-of-phase} density oscillation in both 1D and 2D. Experiments have already shown that a pattern associated with spin density waves can be generated with a modulation of the trapping frequency in an elongated, quasi-1D, condensate \cite{Cominotti}. Here, we have numerically obtained similar results by modulating the scattering lengths, and we have studied the dynamical evolution of the pattern, including the long-time evolution, until nonlinear saturation and randomization.  In a pancake-shaped condensate, the modulation generates patterns that exhibit an $l$-fold rotational symmetry with different numbers of radial nodes.  We have shown that one can reconstruct the spectrum of spin waves by systematically looking at the parametric resonance conditions. We have compared different modulation protocols and characterized the dynamical behavior of the Faraday pattern. 

A natural extension of this work would be to move even closer to the miscible-immiscible transition, in order to study the possible interplay between parametric amplification and dynamic (modulational) instabilities that would spontaneously occur in the immiscible phase once the transition has been overcome~\cite{Sabbatini,SAito}.

Our results may stimulate further theoretical and experimental work on parametric excitations in atomic gases, along the lines of similar studies conducted on Rabi-coupled BECs \cite{Chen}, BEC with dipolar interactions \cite{Nath,Vudragovic,Turmanov}, spinor BEC \cite{SAito}, spin-orbit-coupled BEC \cite{X. Li,Otlaadisa,Luo,Q. Li,Bhat,Zhang2022}, and BEC in optical lattices \cite{Luo,Dalfovo1}. Faraday patterns can be an effective tool for studying spin excitations in multi-component condensates, offering a useful complement to other spectroscopic methods.

\begin{acknowledgments}
M.W., J.W., and Y.L. are supported by National Natural Science Foundation of China (Grant Nos. 11774093, 12074120), Natural Science Foundation of Shanghai (Grant No. 23ZR1418700), Innovation Program of Shanghai Municipal Education Commission (Grant No. 202101070008E00099) and Program of Chongqing Natural Science Foundation (Grant No. CSTB2022NSCQ-MSX0585). C.Q. is supported by ACC-New Jersey under Contract No. W15QKN-18-D-0040. F.D. is supported by Provincia autonoma di Trento. 
\end{acknowledgments}



\begin{thebibliography}{99}

\bibitem{Cross} M. C. Cross, and P. C. Hohenberg, \textit{Pattern formation outside of equilibrium}, Rev. Mod. Phys. \textbf{65}, 851 (1993).

\bibitem{Arecchi} F. Arecchi, S. Boccaletti, and P. Ramazza, \textit{Pattern formation and competition in nonlinear optics}, Phys. Rep. \textbf{318}, 1 (1999).

\bibitem{Liddle} A. R. Liddle, and D. H. Lyth, \textit{Cosmological inflation and large-scale structure}, Cambridge University Press (2000).

\bibitem{Faraday} M. Faraday, \textit{XVII. On a peculiar class of acoustical figures; and on certain forms assumed by groups of particles upon vibrating elastic surfaces}, Phil. Trans. R. Soc. \textbf{121}, 299 (1831).

\bibitem{Douady} S. Douady, \textit{Experimental study of the Faraday instability}, J. Fluid Mech. \textbf{221}, 383 (1990).

\bibitem{Westra} M. T. Westra, D. J. Binks, and Willem van de Water, \textit{Patterns of Faraday waves}, J. Fluid Mech. \textbf{496}, 1 (2003).

\bibitem{Zhao} X. Zhao, J. Tang, and J. Liu, \textit{Electrically switchable surface waves and bouncing droplets excited on a liquid metal bath}, Phys. Rev. F \textbf{3}, 124804 (2018).

\bibitem{Maini} P. K. Maini, K. J. Painter, and H. N. P. Chau, \textit{Spatial pattern formation in chemical and biological systems}, J. Chem. Soc., Faraday Trans. \textbf{93}, 3601 (1997).

\bibitem{Petrov} V. Petrov, Q. Ouyang, and H. L. Swinney, \textit{Resonant pattern
formation in a chemical system}, Nature (London) \textbf{388}, 655 (1997).

\bibitem{Kovacic} I. Kovacic, R. Rand, and S. Mohamed Sah, \textit{Mathieu's Equation and Its Generalizations: Overview of Stability Charts and Their Features}, Appl. Mech. Rev. \textbf{70}, 020802 (2018).

\bibitem{Barone} S. R. Barone, M. A. Narcowich, and F. J. Narcowich, \textit{Floquet theory and applications}, Phys. Rev. A \textbf{15}, 1109 (1977).

\bibitem{pitaevskii-16} L. P. Pitaevskii, and S. Stringari,  \textit{Bose-Einstein Condensation and Superfluidity}, Oxford University Press, UK, (2016).

\bibitem{Dalfovo1} M. Kramer, C. Tozzo, and F. Dalfovo, \textit{Parametric excitation of a Bose-Einstein condensate in a one-dimensional optical lattice}, Phys. Rev. A \textbf{71}, 061602 (R) (2005).

\bibitem{Dalfovo2} C. Tozzo, M. Kramer, and F. Dalfovo, \textit{Stability diagram and growth rate of parametric resonance in Bose-Einstein condensates in one-dimensional optical lattices}, Phys. Rev. A \textbf{72}, 023613 (2005).

\bibitem{Dalfovo3} M. Modugno, C. Tozzo, and F. Dalfovo, \textit{Detecting phonons and persistent currents in toroidal Bose-Einstein condensates by means of pattern formation}, Phys. Rev. A \textbf{74}, 061601(R) (2006).

\bibitem{Nicolin} A. I. Nicolin, R. C.-Gonzalez, and P. G. Kevrekidis, \textit{Faraday waves in Bose-Einstein condensates}, Phys. Rev. A \textbf{76}, 063609 (2007).

\bibitem{Engels} P. Engels, C. Atherton, and M. A. Hoefer, \textit{Observation of Faraday Waves in a Bose-Einstein Condensate}, Phys. Rev. Lett. \textbf{98}, 095301 (2007).

\bibitem{Nicolin2} A. I. Nicolin, \textit{Resonant wave formation in Bose-Einstein condensates}, Phys. Rev. E \textbf{84}, 056202 (2011).

\bibitem{Jaskula} J.-C. Jaskula, G. B. Partridge, M. Bonneau, R. Lopes, J. Ruaudel, D. Boiron, and C. I. Westbrook, \textit{Acoustic Analog to the Dynamical Casimir Effect in a Bose-Einstein Condensate}, Phys. Rev. Lett. \textbf{109}, 220401 (2012).

\bibitem{Smits} J. Smits, L. Liao, H. T. C. Stoof, and P. van der Straten, \textit{Observation of a Space-Time Crystal in a Superfluid Quantum Gas}, Phys. Rev. Lett. \textbf{121}, 185301 (2018).

\bibitem{Brazhnyi} V. A. Brazhnyi, and V. V. Konotop, \textit{Theory of nonlinear matter waves in optical lattices}, Mod. Phys. Lett. B \textbf{18}, 627 (2004).

\bibitem{Kartashov} Y. V. Kartashov, B. A. Malomed, and L. Torner, \textit{Solitons in nonlinear lattices}, Rev. Mod. Phys. \textbf{83}, 405 (2011).

\bibitem{Staliunas} K. Staliunas, S. Longhi, and G. J. de Valcárcel, \textit{Faraday Patterns in Bose-Einstein Condensates}, Phys. Rev. Lett. \textbf{89}, 210406 (2002).

\bibitem{Vidanovic} I. Vidanović, A. Balaž, H. Al-Jibbouri, and A. Pelster, \textit{Nonlinear Bose-Einstein-condensate dynamics induced by a harmonic modulation of the $s$-wave scattering length}, Phys. Rev. A \textbf{84}, 013618 (2011).

\bibitem{Okazaki} K. Okazaki, J. Han, and M. Tsubota, \textit{Faraday waves in Bose--Einstein condensate: From instability to nonlinear dynamics}, arXiv:2012.02391.

\bibitem{Nguyen} J. H. V. Nguyen, M. C. Tsatsos, D. Luo, A. U. J. Lode, G. D. Telles, V. S. Bagnato, and R. G. Hulet, \textit{Parametric Excitation of a Bose-Einstein Condensate: From Faraday Waves to Granulation}, Phys. Rev. X \textbf{9}, 011052 (2019).

\bibitem{Kwon} K. Kwon, K. Mukherjee, S. J. Huh, K. Kim, S. I. Mistakidis, D. K. Maity, P. G. Kevrekidis, S. Majumder, P. Schmelcher, and J. Y. Choi, \textit{Spontaneous Formation of Star-Shaped Surface Patterns in a Driven Bose-Einstein Condensate}, Phys. Rev. Lett. \textbf{127}, 113001 (2021).

\bibitem{Maity} D. K. Maity, K. Mukherjee, S. I. Mistakidis, S. Das, P. G.
Kevrekidis, S. Majumder, and P. Schmelcher, \textit{Parametrically excited star-shaped patterns at the interface of binary Bose-Einstein condensates}, Phys. Rev. A \textbf{102}, 033320 (2020).

\bibitem{Zhang} Z. Zhang, K.-X. Yao, L. Feng, J. Hu, and C. Chin, \textit{Pattern formation in a driven Bose–Einstein condensate}, Nat. Phys. \textbf{16}, 652 (2020).

\bibitem{Rajkov} D. Hern\'{a}ndez-Rajkov, J. E. Padilla-Castillo, A. del R\'{i}o-Lima, A. Guti\'{e}rrez-Vald\'{e}s, F. J. Poveda-Cuevas, and J. A. Seman, \textit{Faraday waves in strongly interacting superfluids}, New J. Phys. \textbf{23}, 103038 (2021).

\bibitem{Recati} A. Recati, and S. Stringari, \textit{Coherently Coupled Mixtures of Ultracold Atomic Gases}, Annu. Rev. Condens. Matter Phys. \textbf{13}, 407 (2022).

\bibitem{Baroni} C. Baroni, G. Lamporesi, and M. Zaccanti, \textit{Quantum mixtures of ultracold gases of neutral atoms}, Nat. Rev. Phys. \textbf{6}, 736-752 (2024). 

\bibitem{Papp} S. B. Papp, J. M. Pino, and C. E. Wieman, \textit{Tunable miscibility in a dual-species Bose-Einstein condensate}, Phys. Rev. Lett. \textbf{101}, 040402 (2008).

\bibitem{Wacker} L. Wacker, N. B. Jorgensen, D. Birkmose, R. Horchani, W. Ertmer, C. Klempt, N. Winter, J. Sherson, and J. J. Arlt, \textit{Tunable dual-species Bose-Einstein condensates of $^{39}$K and $^{87}$Rb}, Phys. Rev. A \textbf{92}, 053602 (2015).

\bibitem{Matthews} M. R. Matthews, B. P. Anderson, P. C. Haljan, D. S. Hall, M. J. Holland, J. E. Williams, C. E. Wieman, and E. A. Cornell, \textit{Watching a Superfluid Untwist Itself: Recurrence of Rabi Oscillations in a Bose-Einstein Condensate}, Phys. Rev. Lett. \textbf{83}, 3358 (1999).

\bibitem{Lin} Y.-J. Lin, K. Jim\'{e}nez-Garc\'{i}a, and I. B. Spielman, \textit{Spin–orbit-coupled Bose–Einstein condensates}, Nature \textbf{471}, 83-86 (2011). 

\bibitem{Sabbatini} J.~Sabbatini, W.H.~Zurek and M.J.~Davis, \textit{Causality and defect formation in the dynamics of an engineered quantum phase transition in a coupled binary Bose–Einstein condensate}, New J. Phys. \textbf{14}, 095030 (2012).

\bibitem{Cominotti} R. Cominotti, A. Berti, A. Farolfi, A. Zenesini, G.
Lamporesi, I. Carusotto, A. Recati, and G. Ferrari, \textit{Observation of Massless and Massive Collective Excitations with Faraday Patterns in a Two-Component Superfluid}, Phys. Rev. Lett. \textbf{128}, 210401 (2022).

\bibitem{Chai} X. Chai, D. Lao, K. Fujimoto, R. Hamazaki, M. Ueda,
and C. Raman, \textit{Magnetic Solitons in a Spin-1 Bose-Einstein Condensate}, Phys. Rev. Lett. \textbf{125}, 030402 (2020).

\bibitem{Farolfi}  A. Farolfi, D. Trypogeorgos, C. Mordini, G. Lamporesi,
and G. Ferrari, \textit{Observation of Magnetic Solitons in Two-Component Bose-Einstein Condensate}, Phys. Rev. Lett. \textbf{125}, 030401 (2020).

\bibitem{Qu} C. Qu, L. P. Pitaevskii, and S. Stringari, \textit{Magnetic Solitons in a Binary Bose-Einstein Condensate}, Phys. Rev. Lett. \textbf{116}, 160402 (2016).

\bibitem{Bienaimé}  T. Bienaimé, E. Fava, G. Colzi, C. Mordini, S. Serafini, C.
Qu, S. Stringari, G. Lamporesi, and G. Ferrari, \textit{Spin-dipole
oscillation and polarizability of a binary Bose-Einstein
condensate near the miscible-immiscible phase transition}, Phys. Rev. A
\textbf{94}, 063652 (2016).

\bibitem{Kim}J. H. Kim, D. Hong, and Y. Shin, \textit{Observation of two sound modes in a binary superfluid gas}, Phys. Rev. A \textbf{101}, 061601 (2020).

\bibitem{Pu} H. Pu and N. P. Bigelow, \textit{Properties of Two-Species Bose Condensates}, Phys. Rev. Lett. \textbf{80}, 1130 (1998); \textit{Collective Excitations, Metastability, and Nonlinear Response of a Trapped Two-Species Bose-Einstein Condensate}, Phys. Rev. Lett. \textbf{80}, 1134 (1998).

\bibitem{Timmermans} E. Timmermans, \textit{Phase Separation of Bose-Einstein Condensates}, Phys. Rev. Lett. \textbf{81}, 5718 (1998).

\bibitem{Pethick} C. J. Pethick, and H. Smith, \textit{Bose-Einstein Condensation in dilute Gases}
(Cambridge University Press,
Cambridge, England, 2002).

\bibitem{E.Taylor} E. Taylor, and E. Zaremba, \textit{Bogoliubov sound speed in periodically modulated Bose-Einstein condensates}, Phys. Rev. A \textbf{68}, 053611 (2003).

\bibitem{Kramer} M. Kr$\ddot{\mathrm{a}  }$mer, C. Menotti, and M. Modugno, \textit{Velocity of sound in a Bose-Einstein condensate in the presence of an optical lattice and transverse confinement}, J. Low Temp. Phys. \textbf{138}, 729-734 (2005).

\bibitem{Ticknor1} C. Ticknor, \textit{Excitations of a trapped two-component Bose-Einstein condensate}, Phys. Rev. A \textbf{88},103623 (2013).

\bibitem{Ticknor2} C. Ticknor, \textit{Dispersion relation and excitation character of a two-component Bose-Einstein condensate}, Phys. Rev. A \textbf{89}, 053601 (2014).

\bibitem{Minniti} F. Dalfovo, C. Minniti, and L.P. Pitaevskii, \textit{Frequency shift and mode coupling in the nonlinear dynamics of a {B}ose-condensed gas}, Phys. Rev. A \textbf{56}, 4855 (1997).

\bibitem{Liebster} N. Liebster, M. Sparn, E. Kath, J. Duchene, K. Fujii, S.L. G\"orlitz, T. Enss, H. Strobel, and M.K. Oberthaler, \textit{Observation of Pattern Stabilization in a Driven Superfluid}, Phys. Rev. X \textbf{15}, 011026 (2025).

\bibitem{SAito} H. Saito, and M. Ueda, \textit{Spontaneous magnetization and structure formation in a spin-1 ferromagnetic Bose-Einstein condensate}, Phys. Rev. A \textbf{72}, 023610 (2005).

\bibitem{Chen} T. Chen, K. Shibata, Y. Eto, T. Hirano, and H. Saito, \textit{Faraday patterns generated by Rabi oscillation in a binary
Bose-Einstein condensate,}
 Phys. Rev. A \textbf{100}, 063610 (2019).

\bibitem{Nath} R. Nath, and L. Santos, \textit{Faraday patterns in two-dimensional dipolar Bose-Einstein condensates}, Phys. Rev. A \textbf{81}, 033626 (2010).

\bibitem{Vudragovic} D. Vudragovic, and A. Balaz, \textit{Faraday and Resonant Waves in Dipolar Cigar-Shaped Bose-Einstein Condensates}, Symmetry \textbf{11}, 1090 (2019).

\bibitem{Turmanov} B. K. Turmanov, B. B. Baizakov, and F. K. Abdullaev, \textit{Generation of density waves in dipolar quantum gases by time-periodic modulation of atomic interactions}, Phys. Rev. A \textbf{101}, 053616 (2020).

\bibitem{X. Li} X.-X. Li, R.-J. Cheng, A.-X. Zhang, and J.-K. Xue, \textit{Modulational instability of Bose-Einstein condensates with helicoidal spin-orbit coupling}, Phys. Rev. E \textbf{100}, 032220 (2019).

\bibitem{Otlaadisa} P. Otlaadisa, C. B. Tabi, and T. C. Kofan\'{e}, \textit{Modulation instability in helicoidal spin-orbit coupled open Bose-Einstein condensates}, Phys. Rev. E \textbf{103}, 052206 (2021).

\bibitem{Luo} X. Luo, B. Yang, J. Cui, Y. Guo, L. Li, and Q. Hu, \textit{Dynamics of spin–orbit-coupled cold atomic gases in a Floquet lattice with an impurity}, J. Phys. B \textbf{52}, 085301 (2019).

\bibitem{Q. Li} G.-Q. Li, G.-D. Chen, P. Peng, Z. Li, and X.-D. Bai, \textit{Modulation instability of a spin-1 Bose-Einstein condensate with spin-orbit coupling}, J. Phys. B \textbf{50}, 235302 (2017).

\bibitem{Bhat} I. A. Bhat, T. Mithun, B. A. Malomed, and K. Porsezian, \textit{Modulational instability in binary spin-orbit-coupled Bose-Einstein condensates}, Phys. Rev. A \textbf{92}, 063606 (2015).

\bibitem{Zhang2022} H. Zhang, S. Liu, and Y. Zhang, \textit{Faraday patterns in spin-orbit-coupled Bose-Einstein condensates,}
 Phys. Rev. A \textbf{105}, 063319 (2022).



\end{thebibliography}
\end{document}